\documentclass[aps,prl,reprint,bibnotes,footinbib,showkeys,groupedaddress,superscriptaddress]{revtex4-1}
\pdfoutput=1
\usepackage{graphicx}
\usepackage{amsmath,amsfonts,amssymb,amsthm}
\usepackage{verbatim}

\usepackage{natbib}
\usepackage{color}
\definecolor{darkblue}{rgb}{0,0,0.5}
\definecolor{darkred}{rgb}{0.5,0,0}
\usepackage[colorlinks=true,urlcolor=darkblue,citecolor=darkblue,linkcolor=darkred,hyperfootnotes=false]{hyperref}
\usepackage{mathbbol}

\usepackage{amsmath,amsfonts,amssymb,amsthm}
\usepackage{graphicx}% Include figure files
\usepackage{dcolumn}% Align table columns on decimal point
\usepackage{bm}% bold math
\usepackage{siunitx}

\usepackage[normalem]{ulem}

\newcommand{\bo}[1]{\mathbf{#1}}

\newcommand{\etal}{\textit{et al}.~}

\begin{document}

\title{Run-to-tumble variability controls the surface residence times of {\it E.~coli} bacteria}

\author{Gaspard Junot$^1$, Thierry Darnige$^1$, Anke Lindner$^1$, Vincent A. Martinez$^3$ \\ Jochen Arlt$^3$, Angela Dawson$^3$, Wilson C. K. Poon$^3$, Harold Auradou$^2$, Eric Cl\'ement$^{4,}$}

\affiliation{Laboratoire PMMH-ESPCI Paris, PSL Research University, Sorbonne Universit\'e and Denis Diderot, 7, quai Saint-Bernard, Paris, France.\\$^{2}$
Universit\'{e} Paris-Saclay, CNRS, FAST, 91405, Orsay, France.
\\$^{3}$SUPA and the School of Physics \& Astronomy, The University of Edinburgh, Peter Guthrie Tait Road, Edinburgh EH9 3FD, United Kingdom.\\ $^{4}$Institut Universitaire de France (IUF).}
\date{\today}

\begin{abstract}
Motile bacteria are known to accumulate at surfaces, eventually leading to changes in bacterial motility and bio-film formation. We use a novel two-colour, three-dimensional Lagrangian tracking technique, to follow simultaneously the body and the flagella of a wild-type {\it Escherichia~coli}. We observe long surface residence times and surface escape corresponding mostly to immediately antecedent tumbling. A motility model accounting for a large behavioural variability in run-time duration, reproduces all experimental findings and gives new insights into surface trapping efficiency.
\end{abstract}

% PACS,

\maketitle

Suspensions of active particles such as motile microorganisms display rich, often counter-intuitive, phenomena unseen in suspensions of passive colloids \cite{ReviewActiveMatter2013} as for example, effective viscosity lower than the pure solvent \cite{Matias2015,Martinez2020}, formation of "living crystals" \cite{Petroff2015} or accumulation at the walls \cite{Berke2008,li2009accumulation,Drescher}. Persistence in the swimming direction along surfaces is a generic contributing to "surface trapping" along with hydrodynamic \cite{Berke2008} or eventually transient adhesion \cite{Perez2019}.
Bacterial surface motility is involved in many industrial, biomedical or environmental issues, such as bacterial contamination or bio-fouling \cite{Schultz2011,Bixler2012}. 
Attachment of bacteria to surfaces often leads to the build-up of hard-to-eradicate bio-films and is problematic for medical implants \cite{Bruellhoff2010}, water purification systems \cite{Kang2012} and many industrial processes \cite{Marcato2012}. 
In nature, the attachment of bacteria to plant roots constitutes the first physical step in many plant–microbe interactions\cite{Wheatley2018}.
Adhesion may originate from surface restriction to flagellar motion \cite{Petrova2012} and trigger the secretion of polysaccharides for structuring mature bio-films \cite{Tuson2013}. The initial stage preceding surface adhesion is therefore the `residence time' $\tau$ of the swimming bacterium at the surface. This quantity is key to understand and model the problems of bacterial contamination in environmental or bio-medical situations \cite{creppy2019effect,figueroa2020coli}.
Wild-type (WT) {\it Escherichia coli} perform run and tumble (R\&T), in which straight runs are interspersed with tumbles where the swimming direction changes rapidly. 
The escape mechanism of those bacteria is still not fully understood \cite{mears2014escherichia,turner2016visualizing}. To date, the detailed micro-hydrodynamics of this phenomenon remains challenging even for state-of-the-art numerics \cite{eisenstecken2016bacterial}. Recent experiments using digital holographic microscopy to capture 3D trajectories of wild-type {\it E.~coli} \cite{molaei2014failed} near a solid surface, suggested that surfaces inhibit tumbling and polarise the post-tumbling direction parallel to the surface, so that tumbling is not a particularly effective escaping mechanism.\\ 
\begin{figure*}[ht!]
\centering
	\includegraphics[width= 18cm]{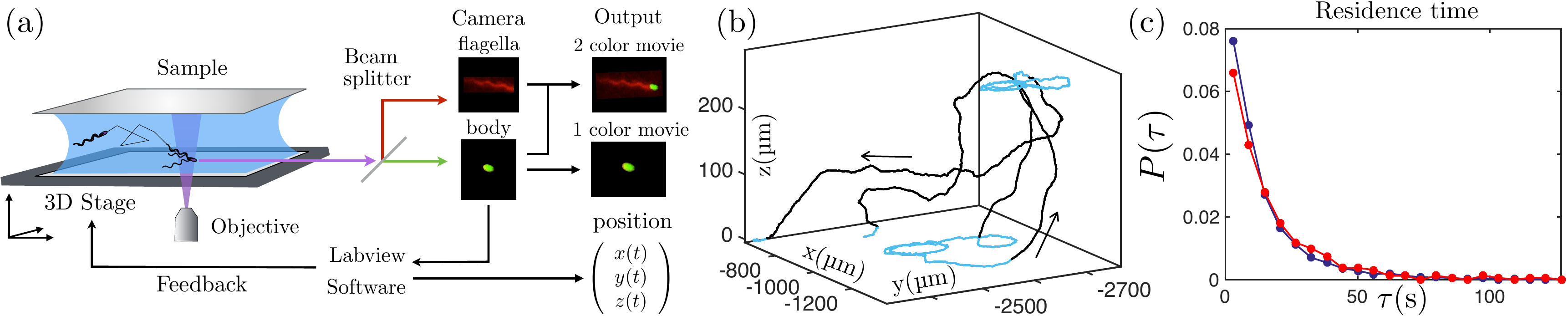}
	\vspace{0 mm}
	\caption{(a) Sketch of the Lagrangian tracking giving the position of the bacterium and videos in one or two colours. (b) 3D view of a bacterium trajectory, parts of the trajectory below \SI{8}{\micro\meter} from the surface are in blue, arrows indicate the trajectory direction. (c) Distribution $P(\tau)$ of residence times $\tau$ at the surface. Experimental data are in red and simulations in blue.}
\label{Fig_setup}
\end{figure*}
In our study, individual motile {\it E.~coli} bacteria were tracked using two-colour three-dimensional tracking (2C-3DT) that provides visualisation of the flagella dynamics with an unprecedented precision. 
Observations of displacements close to a surface were made during long periods of times allowing, for the first time, the assessment of the surface residence time distribution, the angular distributions for arrival and escape and the distribution of duration of unbundling events.
Those distributions were compared with measurements performed away from the surface. We find long surface residence times and demonstrate that tumbling is the dominant escape mechanism. To reproduce our observations, we adapted and simulated a recent 'behavioural variability' (BV) model \cite{figueroa20203d} in which the run-to-tumble transition displays a much larger variability compared to a Poisson probability distribution of transition \cite{Berg2004}.

{\it Methods---} We implemented 2C-3DT by combining Lagrangian 3D tracking \cite{Darnige2016,figueroa20203d} with two-colour fluorescence imaging \cite{schwarz-linek2016escherichia} (See Fig.~\ref {Fig_setup}(a)) on an inverted epifluorescence microscope (Zeiss-Observer, Z1, C-Apochromat 63$\times$/1.2~W objective). To avoid signal overlap and emission leakage, we engineered an {\it E.~coli} strain (AD62) with body and flagella fluorescence in the green and red respectively (see SI). A two-colour LED light source (Zeiss Colibri 7) and a dichroic image splitter (Hamamatsu) are used to project two monochrome images onto two different regions of the camera chip. Computer-controlled movement of the microscope stage keeps the body of a selected bacterium in focus \cite{Darnige2016} and images (1024$\times$1024 pixels) are recorded at 80 fps with an Hamamatsu  ORCAFlash  4.0,C11440 camera. Green and red images are then superimposed to create a movie of the tracked bacterium and its flagella bundle (see video SI). 
Photo-bleaching limits flagella imaging to a minute and thus, long time behaviour can not be observed with this technique. For long-time tracking, we use a strain with non-fluorescent flagella (RP437) that allows one-colour recording 
of 66 independent cells over \SI{7}{\hour}, with the longest track being of $\gtrsim \SI{20}{\minute}$ duration. 
Bacteria were grown and prepared using standard protocols \cite{schwarz-linek2016escherichia} (see SI). For imaging, a \SI{80}{\micro\liter} drop with $\lesssim 3\times 10^7$ cells \si{\per\milli\liter} was placed between two glass plates separated by \SI{260}{\micro\meter} and sealed. \\

\begin{figure}[t]
\centering
	\includegraphics[width= 8cm]{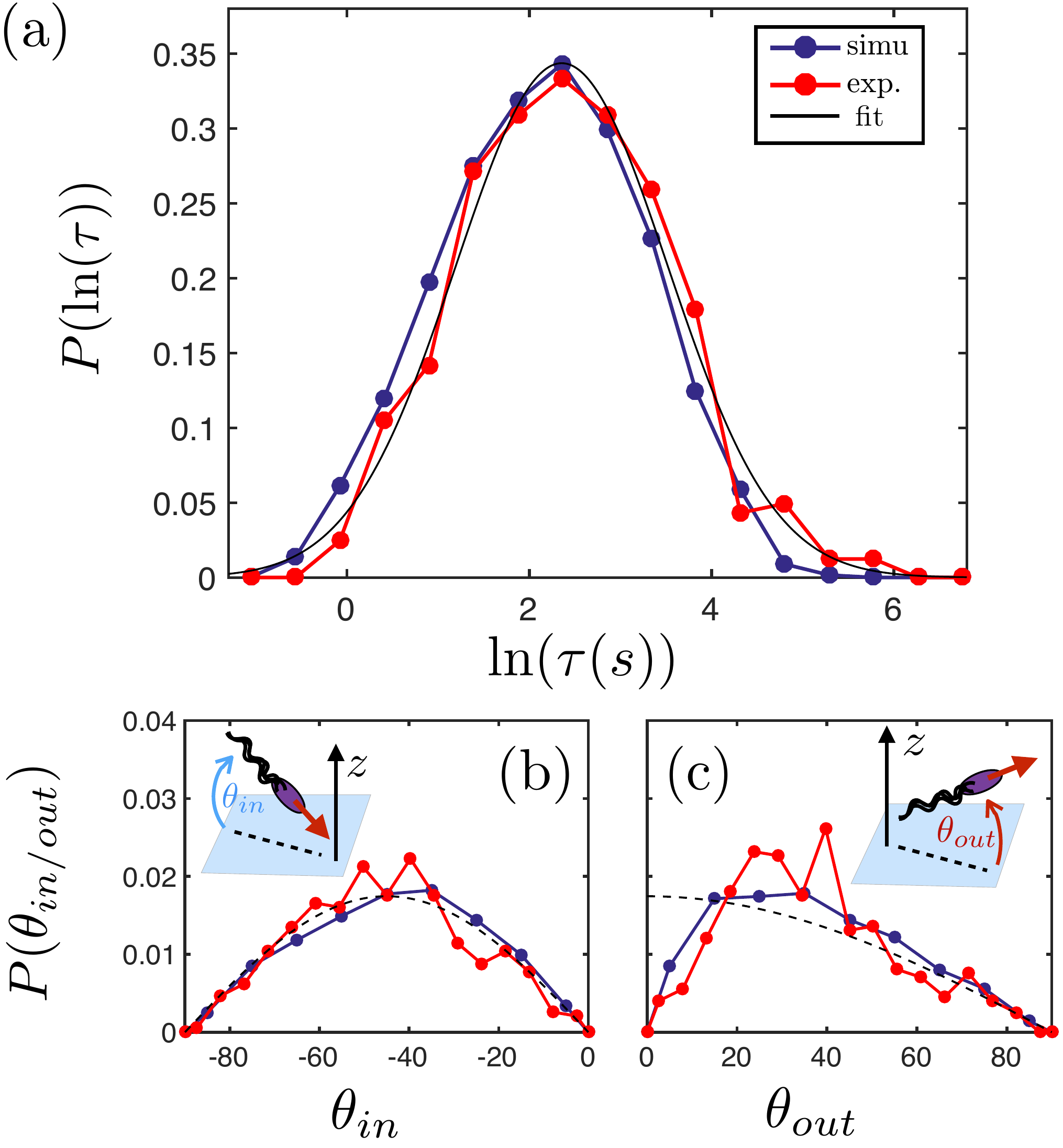}
	\vspace{0 mm}
\caption{Comparison between experiment (in red) and simulations (in blue). (a) Distribution of the logarithm of the residence time ($<\ln(\tau)>=2.39$ and $\sigma = 1.12$). Experimental data corresponds to Fig. \ref{Fig_setup}(c), the black line is a Gaussian fit ($<\ln(\tau)>=2.36$ and $\sigma = 1.16$). (b-c) Distributions of the incoming and escape angles from ``single-color'' tracking. The black dashed line is the distribution $P(\theta_{\rm in})=-\frac{\pi}{180}\sin(\theta_{\rm in})\cos(\theta_{\rm in})$ in (b) and $P(\theta_{\rm out})=\frac{\pi}{180}\cos(\theta_{\rm out})$ in (c).}
\label{Fig_exp_simu}
\end{figure}

\begin{figure}[t!]
\centering
	\includegraphics[width= 8cm]{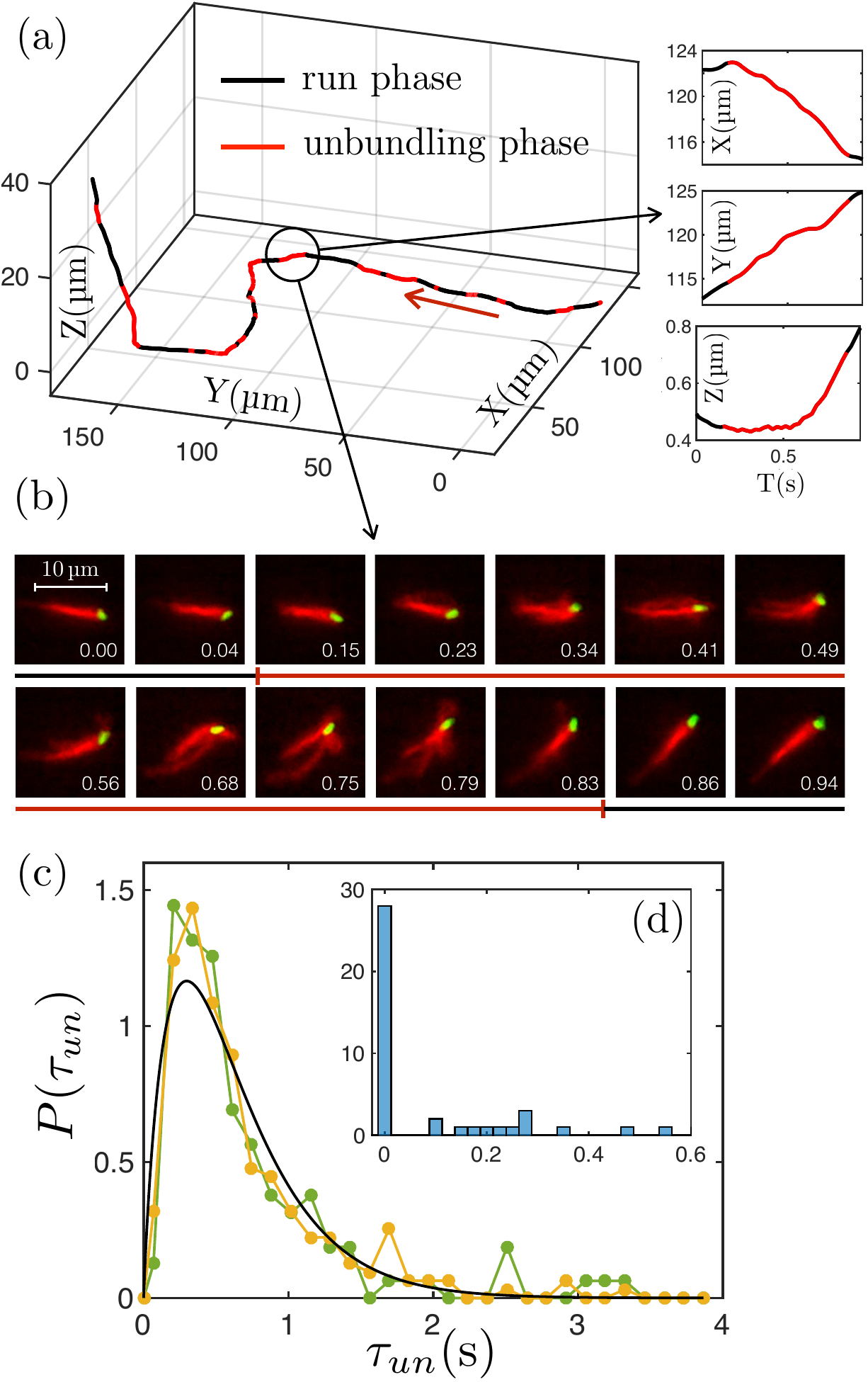}
	\vspace{0 mm}
\caption{(a) 3D trajectory and $x(t),y(t),z(t)$ coordinates of a bacterium (AD62) near the surface. Red parts show unbundling events and the arrow the trajectory direction. (b) Time lapse of an unbundling event, each image is an overlay of 3 consecutive frames. The colour of the line indicates when the unbundling event starts and ends, the total duration is \SI{0.71}{\second}. (c) Distribution of the unbundled time $\tau_{un}$ at surfaces (orange) and in the bulk (green). The black line is a fit using a gamma distribution of parameter $(k,\theta)= (1.9,\SI{0.33}{\second}) $. (d) histogram of time $\tau_d$ between an escape event and the closest previous unbundling event.}
\label{Fig_tumble}
\end{figure}

{\it Experimental results---}To measure the surface residence time, $\tau$, Fig.~\ref{Fig_setup}(b), we need to identify when a cell arrives and escapes from a surface. A bacterium is considered in the bulk when the body centroid is at a distance from the nearest surface $\Delta z > \delta =\SI{8}{\micro\meter}$ (a typical cell body + flagella length) and arrived at a surface when $\Delta z< \SI{3}{\micro\meter}$. The "surface region" is left again when subsequently, $\Delta z > \delta$. The residence time $\tau$ is then the interval between the first and last time a bacterium crosses $\Delta z=\SI{3}{\micro\meter}$ and is not influenced by small variations in the choice of these two lengths (see SI).
The measured distribution of surface residence times $P(\tau)$, Fig. \ref{Fig_setup}(c), has mean $\langle \tau \rangle=\SI{21}{\second}$ and a long tail extending to a maximum observed $\tau$ of $\SI{373}{\second} \lesssim 20 \tau$. 
The long-tailed, highly-non-exponential, nature of $P(\tau)$ is emphasised when plotted against $\ln \tau$, Fig.~\ref{Fig_exp_simu}(a) and fitted to a log-normal distribution.
These residence times are very long as compared to the -usually reported- average run time of WT {\it E.~coli} ($\sim \SI{1}{\second}$ according to ref.\cite{Berg1972}). So, a bacterium seems to tumble many times during its residence at a surface before escaping, apparently confirming the suggestion that tumbling would be an inefficient escape mechanism \cite{molaei2014failed}.
For this track series, we also measured the incoming and escape angles for cells arriving ($\theta_{\rm in}$) and leaving ($\theta_{\rm out}$) the surface region, defined as: $\theta_{\rm in,out}=\arcsin (\bo{p_{\rm in,out}} \cdot \bo{n})$ where $\bo{p_{\rm in,out}}$ is a unit vector aligned with the body of the bacterium and $\bo{n}$ a unit vector normal to the surface. The probability distributions, Fig.~\ref{Fig_exp_simu}(b-c), are obtained from 366 pieces of bacterial tracks reaching or leaving the surface.
First, to understand the incoming angle distribution, one can assume a random swimming orientation in the bulk, yielding a probability to have a swimming direction between $\theta$ and $\theta +d\theta$, proportional to $ d(\sin(\theta))=\cos(\theta) d\theta$. For a given time interval, the number of bacteria actually counted, crossing the surface at a distance $\delta$ and heading towards the wall is $\propto V_B \sin(\theta)$ ($V_B$ being the bacterial velocity). Therefore, after normalisation, the probability density to observe bacteria crossing a distance $\delta$ with an angle $\theta_{\rm in}$ is expected to be $P(\theta_{\rm in})= -\frac{\pi}{180} \sin\theta_{\rm in} \cos\theta_{\rm in}$, which agrees with the experimental results (see Fig.~\ref{Fig_exp_simu}(b)).
Now, to understand $P(\theta_{\rm out})$, let us consider the case of a bacterium at a surface with an orientation pointing toward the bulk. If the bacterium does not tumble before reaching the boundary $\Delta z = \delta$, it will cross this height. 
We then expect $P(\theta_{\rm out})= \frac{\pi}{180} \cos\theta_{\rm out}$. Comparing this expression to the experimental data, Fig.\ref{Fig_exp_simu}(c), one can see a dip around $\theta_{\rm out} = 0$ and also a peak around \SI{30}{\degree}. The deficit in the probably density originates from the fact that a cell leaving at a grazing angle ($\theta_{\rm out} \to 0$) needs to swim straight for long times before reaching $\Delta z = \delta$, hence maximising its chances for another tumbling event {\it en route}. This will either reorient the cell back to the surface (failed escape) or the bacterium will be logged at $\Delta z = \delta$, as having escaped at a different (likely higher) angle.\\ 
Next, we characterise the tumbling statistics using 2C-3DT image sequences to identify unambiguously what we call the "unbundling phase", where at least one flagellum is observed outside the flagella bundle. 
Importantly, the "unbundling phase" as defined here does not necessarily mean continuous and uninterrupted changes of direction as already noticed by Turner et al.\cite{turner2016visualizing}. 
These unbundling events are then different from tumbles based on changes of orientation \cite{Berg1972,qu2018changes} or on velocity distributions \cite{Seyrich2018}. In SI, for completeness, we discuss this point extensively. However here, the purpose is to compare characteristic features of the tumbling process in the bulk and at the surface directly issued from the observed flagellar dynamics.
Figure~\ref{Fig_tumble}(a) shows the trajectory of a typical cell swimming at the surface before escaping. We manually identify the beginning and the end of the flagella unbundling process by replaying relevant sequences of the two-colours movie back and forth. A time lapse of a typical unbundling event is shown in Fig.~\ref{Fig_tumble}(b) (see video in SI). From such analysis, we obtain $P(\tau_{un})$, the probability distributions of the bulk and near-wall unbundling phase duration, displayed in Fig.~\ref{Fig_tumble}(c) ($\tau_{un}$ is compiled from 119 and 241 events respectively). 
The two distributions collapse indicating that the surface does not affect the tumbling statistics as visualised on the unbundling events: in each case, $P(\tau_{un})$ is peaked around $\tau_{un}=\SI{0.34}{\second}$ with a mean $\left < \tau_{un} \right > \approx \SI{0.8(1)}{\second}$.
We determine experimentally that only a fraction of $\tau_{un}$ leads to a reorientation and that this fraction can be taken as random within $\tau_{un}$ (see SI).
This would then yields a mean reorientation time of about $0.4 s$, a value significantly larger than the mean tumbling times previously reported (around $0.1 s$ \cite{Berg1972,qu2018changes,Seyrich2018}).
Although tumble events do not always lead to an escape, Fig.~\ref{Fig_tumble}(a), escape is tightly coupled to tumble. The time interval histogram between an escape event and the previous tumble event is narrowly peaked around zero, Fig.~\ref{Fig_tumble}(d) [inset], i.e., almost every escape is immediately preceded by a tumble. In contrast, a smooth-swimmer strain (CR20) with suppressed tumbling, shows residence times longer than our mean observation time of \SI{374}{\second} (see SI). Therefore, tumbling is indeed the dominant escape mechanism for a surface-trapped WT {\it E.~coli} cell.\\

{\it Model and computer simulations---} To understand our experimental findings, we simulate bacterial trajectories using the BV model and parameters taken from Ref. \cite{figueroa20203d} to describe the R\&T statistics. 
This model accounts for an inherent stochasticity due to the concentration fluctuations of a phosphorylated protein, CheY-P, promoting the switching from counterclockwise to clockwise of the flagella motor rotation and initiating the tumbling process \cite{Berg2004,Korobkova2004}. In the model, the internal parameter $\delta X$ represents fluctuations in CheY-P concentration around the mean, normalised by the standard deviation. Its dynamics is modelled by a Ornstein-Uhlenbeck process leading to a tumbling event rate scaling as $\exp[\Delta_n \delta X]$, where parameter $\Delta_n$ is rendering the sensitivity of the run-to-tumble transition to CheY-P concentration (see model details and parameter choices in SI). We call this internal time-resolved variable $\delta X(t)$, the swimmer "mood" since for low values, a bacterium will likely run for a long time and explore large distances whereas for larger values it would rather tumble and locally forage.\\
To model the surfaces, we purposely reduce the complexity of steric hindrance, hydrodynamics and other interactions between a bacterium and a surface \cite{li2009accumulation,Drescher,lauga2006,Berke2008} to simple alignment rules. A particle arriving from the bulk and reaching a surface ($\Delta z=0$) is immediately aligned with it. After tumbling, if the orientation points towards the wall, the cell is re-aligned with the surface keeping $\Delta z=0$. Otherwise, it leaves the surface with this new orientation.
Trajectories simulated using the BV model show a residence time distribution that matches experiments, Fig.~\ref{Fig_exp_simu}(a) without any fitting parameter. 
However, modeling the run time distribution as the uncorrelated Poisson process with an average run time $\approx \SI{1}{\second}$ \cite{Berg2004} does not reproduce the observed $P(\tau)$ as shown in SI. The residence time is then the consequence of the large distribution of run times.
The simulated distributions of $\theta_{\rm in}$ and $\theta_{\rm out}$ also match the experimental observations (see Fig.~\ref{Fig_exp_simu}(b-c)). For $\theta_{\rm in}$, in spite of a tiny but visible deviation with the numerical results, probably rooted in finite confinement effects, one can conclude as Molaie et al.\cite{bianchi2017holographic} that the cell incoming angle is essentially reflecting a random swimming orientation. For $\theta_{\rm out}$, the small-angle `dip' in $P(\theta_{\rm out})$ is reproduced. 
To estimate the extent of this depletion, note that if a bacterium does not reach the escape limit $\Delta z = \delta$ before the mean run time $\langle \tau_r \rangle$ ($=\SI{2.32}{\second}$ in our model), it will likely tumble. The angle corresponding to a travelling time of $\langle \tau_r \rangle$ over a distance $\delta$ at average speed $\bar v =\SI{26}{\micro\meter\per\second}$ is $\frac{\delta}{ \bar v \langle \tau_r \rangle} \approx \SI{7.6}{\degree} $. We therefore expect depletion in $P(\theta_{\rm out})$ at angles $\lesssim \SI{10}{\degree}$, as observed. Note, however, that we do not reproduce numerically the small peak in $P(\theta_{\rm out})$ at $\approx \SI{30}{\degree}$. The excess of escape probability density likely indicates a surface-hindrance effect for high-angle tumbles (also in accordance with ~\cite{molaei2014failed}), which is not included in the model. 
Importantly, the same set of model parameters accounts for observations in the bulk or near a surface leading to the conclusion that, on the time scale of our observations, surfaces do not modify significantly the biochemical circuitry controlling tumbling.\\
\begin{figure}[t]
\centering
	\includegraphics[width= 8cm]{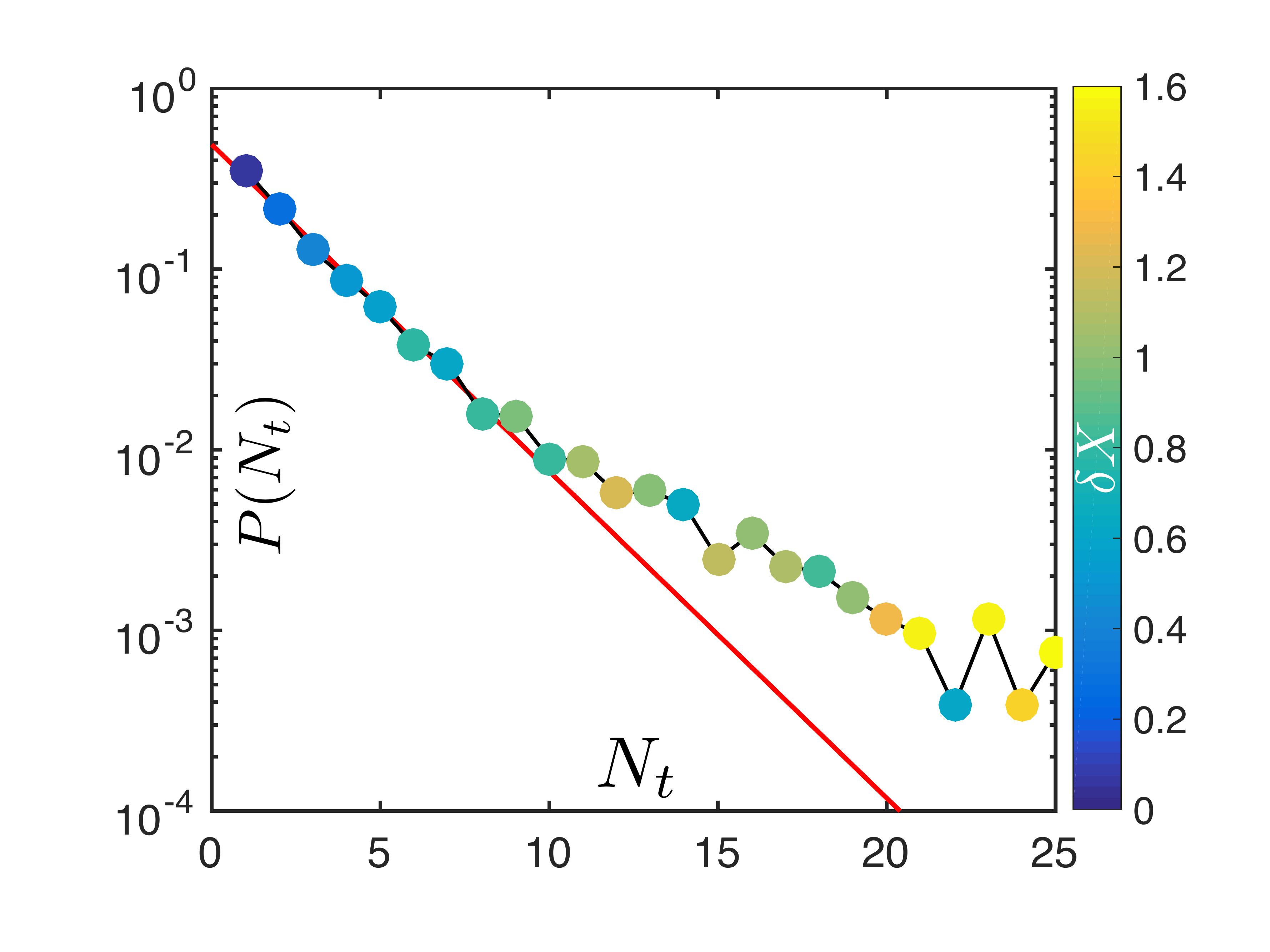}
		\vspace{0 mm}
\caption{Distribution of number of tumbles $N_t$ during a stay at a surface. The red line is proportional to $P(N_t) = (1-p)^{N_t -1}p$ with $p = \frac{1}{2.9}$, which gives a mean number of tumbles of $\langle N_{t} \rangle^\ast=2.4$, compared to $\langle N_t \rangle = 3.5$, the mean of the whole distribution. Symbol colors show the dimensionless average CheY-P concentration $\delta X$ when leaving the surface.}
\label{Fig_Chey}
\end{figure}

From the simulated trajectories, we obtain the probability distribution $P(N_t)$ of the number of tumbles $N_t$ needed for a swimming bacterium trapped in the surface region to finally escape (see Fig.~\ref{Fig_Chey}) 
(Due to flagella bleaching, we could not obtain $P(N_t)$ from the 2C.)
To understand the exponential decay behaviour for $N_t \lesssim 10$, let us consider the probability $p$ to escape out of a single tumbling event, with no memory of the previous tumbling events. Then, the probability to escape after $N_t$ events is $P(N_t) = (1-p)^{N_t -1}p$, or $\log P(N_t) = N_t\log (1-p) + \log\left(\frac{p}{1-p}\right)$. Our data for $N_t \lesssim 10$, Fig.~\ref{Fig_Chey}, are consistent with $p \sim \frac{1}{3}$. 
Noticeably, if all tumbles reorienting a cell away from the surface would lead to a successful escape, we would rather expect $p = \frac{1}{2}$. It is as if post-tumbling re-orientations for angles $< \theta_{\rm min}$ do not lead to escapes, where $\int_{\theta_{\rm min}}^{\pi/2} \frac{\cos\theta}{2} \, \mbox{d}\theta = \frac{1}{3}$, or $\theta_{\rm min} = \arcsin\left(\frac{1}{3}\right) \approx \SI{19}{\degree}$, which is consistent with the extent of the `dip' in the experimental $P(\theta_{\rm out})$ of Fig.~\ref{Fig_exp_simu}(c).
The symbols colours in Fig.~\ref{Fig_Chey} give the mean $\delta X$, hence the ``tumbling mood'' when cells leave the surface region. 
Bluish symbols for $N_t \lesssim 10$ indicate that a majority of escaping bacteria are in a long run-time ``mood'' (or low $\delta X$). These bacteria are likely to escape and go far away from the surface before tumbling again. Thus, they populate the initial exponential decay. In other words, to escape, a bacterium has to tumble while being in a long-run mood.
Yellowish symbols for $N_t \gtrsim 10$ show a minor population of short run-times cells for which tumble does not lead to efficient escape. Their behavior then deviates from the initial exponential decay. 
In both cases, bacteria will stay at surfaces for a long time but for different reasons. 
Numerically, for a mean run-time $<\tau_r>=2.32 s$, we found for bacteria strictly at the surface ($\Delta z=0$) $<\tau_r>=\SI{4.87}{\second}$, for $\Delta z<\delta$, $<\tau_r>= \SI{3.62}{\second}$ and in the bulk, ($\Delta z>\delta$) $<\tau_r>=\SI{1.73}{\second}$. Overall, surfaces act as a preferential selector for longer-run-times in spite of the presence of a frequently tumbling sub-population in the surface region.

{\it Summary and conclusions---} Using a novel 2C-3DT method, we measured for a wild-type {\it E.~coli}, the distributions of residence times at a solid surface, incoming and escaping angles and tumbling times. We found that tumbling is the mechanism by which bacteria escape from surfaces. 
Observations are reproduced quantitatively by a model accounting for a stochasticity in the concentration of a protein (CheY-P) controlling the run to tumble transition rate and leading to a `behavioural variability' of run-times. 
This indicates that the large distribution of residence times is a direct consequence of the non Poissonian run to tumble statistics.
The model solves a paradox where tumbling appears to be a quite efficient mean to escape from surfaces even though wild-type bacteria are likely to be trapped much longer than the typical run time. In this picture, a population of monoclonal bacteria will present a large distribution of motility features, significantly biased by the presence of surfaces. Heterogeneity in bacterial populations is usually seen as the consequence of a variety of selection pressures such as `bet hedging' against environmental change~\cite{Kuipers2011}. Our findings about surface residence prompts the speculation that behavioural variability in the `tumbling mood' may be a form of bet hedging against planktonic and surface living, allowing at every moment different sub-populations to optimise their behaviour relative to chemotaxis in the bulk~\cite{Dev2018} and long surface residence leading to bio-film formation. Evaluating this suggestion obviously requires further research to assess precisely the role of internal noise associated with the chemotactic machinery driving the motor rotation in the context of the different possible `life styles' of {\it E. coli} in their natural habitats. \\ 

%\nocite{dewitt1962occurrence,turner2012growth,merlin2002tools,pilizota2013plasmolysis,Tu2005,saragosti2012modeling}

{\it Acknowledgments.---} This work was supported by grants ``BacFlow'' ANR-15-CE30-0013, CNRS/Royal Society PHC-1576 and IE160675, and the Institut Pierre-Gilles de Gennes (Investissements d'avenir ANR-10-EQPX-34). AL and GJ acknowledge support from the ERC Consolidator Grant PaDyFlow under grant agreement 682367. EC is supported by Institut Universitaire de France. VAM, JA, AD and WCKP were funded by ERC (AdG 340877 PHYSAPS).

\newpage

\section{Supplementary Information}

\subsection{Bacterial culture}
\label{Bacterial culture}
\emph{For strains RP437 and CR20 for which  only the fluorescent body is visualised.}\\
 Bacteria are inoculated in 5mL of culture medium (M9G: 11.3 g/L M9 salt, 4 g/L glucose, 1 g/L casamino acids, 0.1mM CaCl$_2$,  2mM MgSO$_4$) with antibiotics (chloramphenicol at 25$\mu$g/mL for RP437 and amphiciline at 100$\mu$g/mL for CR20) and grown over night at 30$^\circ$C until early stationary phase. The growth medium is then removed by centrifuging the culture and removing the supernatant. The bacteria are re-suspended in a Motility Buffer (MB: 0.1mM EDTA, 0.001mM l-methionine, 10mM sodium lactate, 6.2mM K$_2$HPO$_4$, 3.9mM KH$_2$PO$_4$) with 0.005\%of polyvinyl pyrrolidone (PVP) and is supplemented with 0.08g/mL L-serine. The addition of L-serine increases the bacteria mobility and PVP is classically used to prevent bacteria from sticking to the surfaces. The solution is mixed with Percol (1:1) to avoid bacteria sedimentation. Under these conditions, the average swimming speed is $v_s = 26 \pm 4 \mu$m/s.\\

\emph{For AD62 strains for which the fluorescent body and the flagella are visualised.}\\
Strain AD62 was constructed in E.coli AB1157 \cite{dewitt1962occurrence}.  In AD62 the wild type fliC gene which encodes the flagellin sub-unit  (which polymerises to form the bacterial flagella filament) was modified so that a cysteine was substituted for a serine amino acid at position 219 (S219C).  This allowed labelling of the flagella with Alexa Fluor 647 dye  (Molecular Probes, Life Technologies).  
E.coli HCB1731\cite{turner2012growth} which encodes FliC (S219C) was used as a template to amplify a 803bp fragment of DNA containing the TCA(serine)  to TGC (cysteine) mutation flanked by 400bp of the DNA directly upstream and downstream of this site on each side. Restrictions sites for Xho1 and Sal1 were added at the 5' and 3' ends of this fragment. This fragment of DNA was inserted into plasmid pTOF24 digested with Xho1 and Sal1. The recombinant plasmid was transformed in to E.coli AB1157 and used to replace the wild type version of the gene by allelic exchange as published previously \cite{merlin2002tools}. DNA sequencing confirmed this mutation in the fliC gene.  
This strain was transformed with plasmid pWR21 \cite{pilizota2013plasmolysis}. In E.coli this  plasmid constitutively expresses an eGFP variant to permit fluorescent imaging of the bacterial cell body.\\
\emph{Preparation protocol of  AD62 strains}\\
Suspension of AD62 are prepared using the following protocol: bacteria are inoculated in 10mL of Lurial Broth (LB) with amphiciline at 100$\mu$g/mL and grown over night at 30$^\circ$C. Then 100$\mu$L of this solution is inoculated in 10mL of Triptone Broth (TB) and grown during several hours until early stationary phase. The growth medium is then removed by centrifuging the culture and removing the supernatant. The bacteria are re-suspended in 1mL of Berg Motility Buffer (BMB: 6.2 mM K$_2$HPO$_4$, 3.8 mM $_2$PO$_4$ 4, 67 mM NaCl, and 0.1mM EDTA) with 10$\mu$L of Alexa red colourant (Alexa 647 at 5mg/mL diluted in DMSO) and let under soft shacking during 2 hours. The solution is then washed by centrifuging the culture and removing the supernatant. Finally the bacteria are re-suspended in BMB with 0.005\% of PVP and with 0.08g/mL L-serine.\\
\subsection{Surface definition}
\label{Surface definition}
 
\begin{figure}[ht!]
\centering
	\includegraphics[width= 8cm]{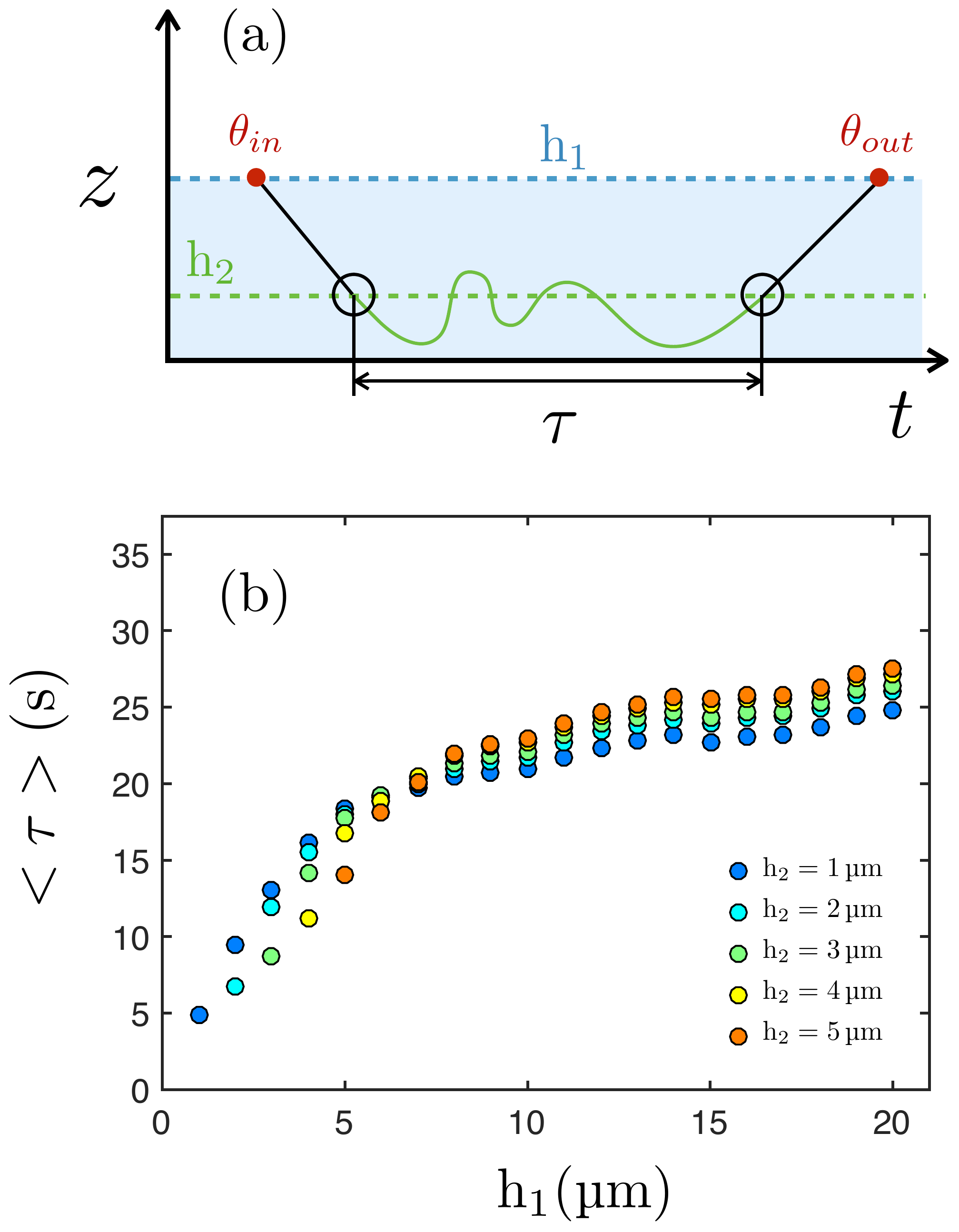}
	\vspace{0 mm}
	\caption{Surface definition with two heights $\text{h}_1$ and $\text{h}_2$. (a) Sketch of a bacterial trajectory, the bacterium enters the surface region, defined below $\text{h}_1$, with an angle $\theta_{in}$ and lives it with an angle $\theta_{out}$. The residence time is computed between the first and last time the bacterium cross $\text{h}_2$. (b) Mean residence time $\langle \tau \rangle$ as function of $\text{h}_1$ and for different $\text{h}_2$.}
	\label{SI_surface_definition}
\end{figure}

We consider as being "in the surface region", portions of trajectories fulfilling two height criteria (see Fig.~\ref{SI_surface_definition}(a)). First, a bacterium coming from the bulk i.e. above a distance $\text{h}_1 = \SI{8}{\micro\meter}$ from the surface (a typical bacterium-length flagella included), has to touch the surface. It means in practice reaching a distance below $\text{h}_2 = \SI{3}{\micro\meter}$ from the surface. 
%This distance is an empirical estimation corresponding to the PDMS surface variations over the whole field of measurements. 
This distance is an empirical estimation corresponding to the uncertainty in the bacterium position relative to the surface. 
Then, to leave the surface it has to cross for the second time, a distance $\text{h}_1$. 
This two heights definition allows small fluctuations in bacterium z-coordinate while swimming at the surface, which will not be the case for a single height definition.
The residence time is then computed between the first and last time the bacterium cross $ \text{h}_2$. We checked that the mean residence time does not strongly depend on the surface definition (see Fig.~\ref{SI_surface_definition}(b)), and we chose $\text{h}_1= \delta = \SI{8}{\micro\meter}$, which is the typical bacterium length (including flagella).

\begin{figure}[ht!]
\centering
	\includegraphics[width= 7cm]{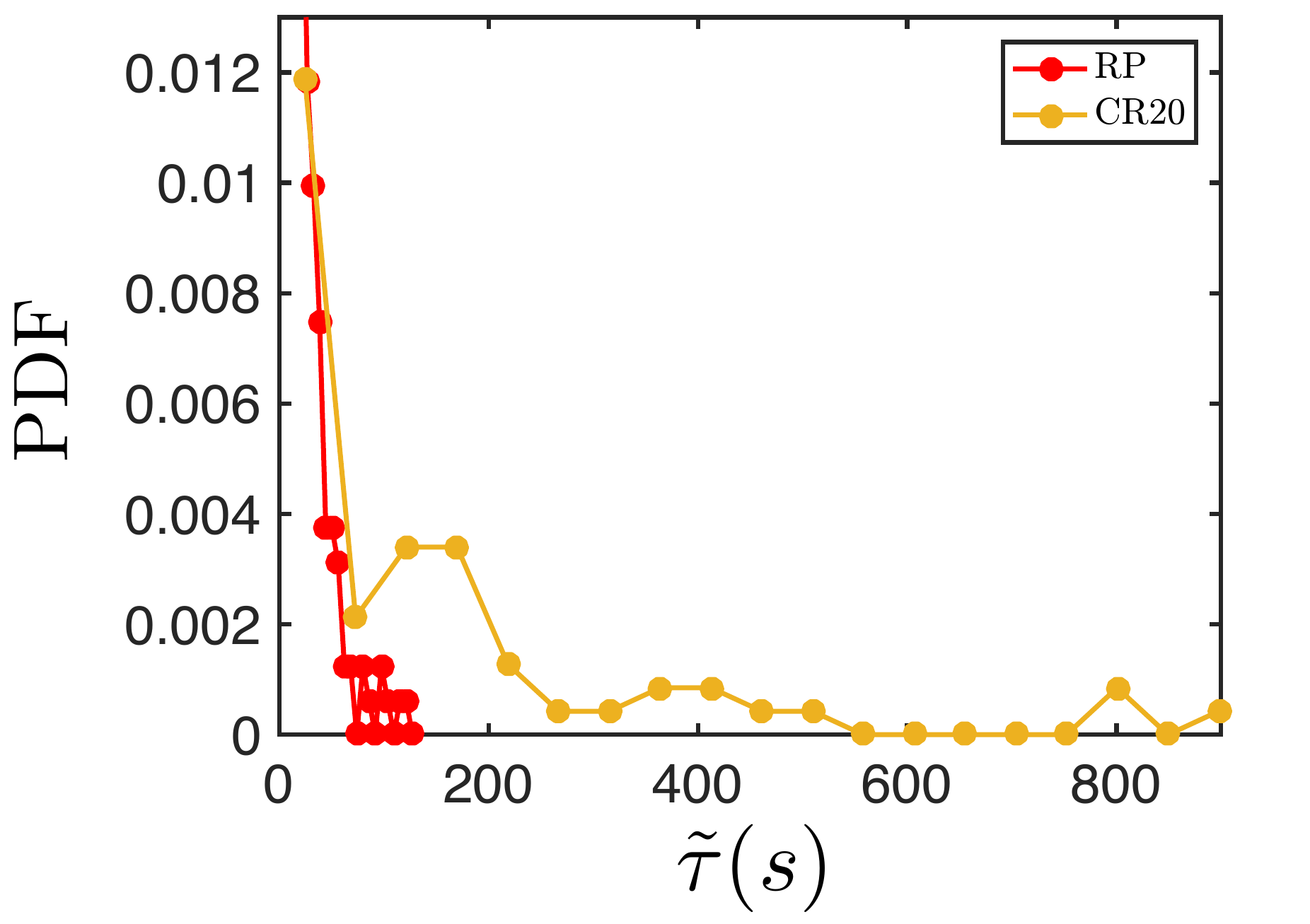}
	\vspace{0 mm}
	\caption{Pseudo residence time distribution of smooth-swimmer CR20 bacteria (orange line) compare to the residence time distribution of the wild-type  RP (red line, same data as in Fig.~\ref{SI_PDF_tau_log_lin}).}
	\label{SI_CR20}
\end{figure}

Smooth-swimmer bacteria (strain CR20), for which tumbles have been inhibited, stay at the surface much longer than the mean tracking time. Therefore, for most of the trajectories, we could not observe the arrival and/or the escape and could not compute a residence time. To be able to compare strains CR20 and RP, we compute a pseudo residence time $\tilde{\tau}$. This time is defined the same way as the residence time $\tau$ except that when the arrival and/or the escape is missing, we take the begin and/or the end of the trajectory. This means that $\tilde{\tau}$ is equal or smaller than $\tau$.
In Fig.~\ref{SI_CR20}, we show the distribution of $\tilde{\tau}$ of the CR20 (orange line) compare to the distribution of $\tau$ of the RP (red line). What we can indeed see is that the smooth-swimmer CR20 stay at the surface much longer that the wild-type RP and that tumbling events are indeed crucial to leave the surface.

 %The mean residence time does not depends strongly on $\text{h}_1$ for $\text{h}_1>\SI{8}{\micro\meter}$ which is the typical bacterium length (including flagella).
 

\subsection{Simulation using the BV model}
\label{Simulation}
\emph{-~General context} \\
Almost 50 years ago Brown and Berg have shown that an ``adapted'' wild-type E.coli, i.e. experiencing no chemical gradient, performs a random walk which is a compound of straight swimming phases (run) and changes of directions (tumble). The tumbling corresponds to a disassembling of the flagella bundle providing the swimming thrust. From the standpoint of the internal chemotactic bio-machinery, tumbling events are associated with phosphorylation of a CheY protein ( CheY + P $\rightarrow$ CheY-P) which changes the direction of rotation of the motors from counter-clockwise rotation (CCW) during the run phase, to clockwise rotation (CW) in the tumbling phase. More recently direct measurements of a sequence of rotation switches on a single cell \cite{Korobkova2004}, demonstrated that the stochastic process associated with the CCW (run phase) turning CW (tumble phase) is not a Poisson process but rather a thick tail distribution. This observation was quantitatively interpreted by Tu and Grinstein \cite{Tu2005} putting forwards a stochastic process based on the concentration fluctuations of the CheY-P protein modifying the energy barrier associated with the phosphorylation chemical process. Recently Figueroa \etal \cite{figueroa20203d} tracked for very long times wild-type E.coli bacteria and showed that the swimming direction persistence displays large behavioural variability with a memory scale around $19 s$. To interpret these experimental measurements, they provided a kinetic model describing the alternation of run and tumble events as well as changes of orientation during the tumbling phase. This model describing the "behavioural variability" of the random walk exploration process was called the ``BV-model''.\\

\emph{-~Description of the BV model} \\
%In their work, they found that the run time does not follow a Poisson distribution, as in the standard model, but rather a log-normal distribution. The switching time to go from the run state to the tumbling state depends on the fluctuation in concentration of a protein. Therefore, depending on the concentration of the protein, the bacterium will either run for a long time or tumble often, justifying the denomination of 
%For wild-type E.coli, the direction of rotation of the rotary motor is triggered by the phosphorilation of a protein named CheY, into CheY-P \cite{bourret2002molecular}. When all the motors turn in the counter-clockwise (CCW) direction all the flagella bundle into an helix and the bacterium runs. Then, when at least one motor turns in the clockwise (CW) direction, the associated flagellum goes outside of the main bundle and the bacterium undergo a reorientation process i.e a tumble.
The key idea of the BV model is that the switching time triggering a run to tumble event depends on the concentration of phosphorylated protein CheY-P. In the model, each bacterium possesses an internal variable  $\delta X$ which represents the relative concentrations of Chey-P around a mean. Following Tu and Grinstein \cite{Tu2005}, the CheY-P concentration $Y$ fluctuates in the motor vicinity as an Ornstein-Uhlenbeck process:
\begin{equation}\label{dynamics_CheY}
\begin{aligned}
\frac{d \delta X}{dt} = -\delta X/T_Y + \sqrt{2/T_Y}\xi (t)
\end{aligned}
\end{equation}
where $\delta X = (\lbrack Y \rbrack-Y_0)/\sigma_Y$ is the normalised concentration of CheY-P with $Y_0$ the mean concentration, $\sigma_Y$ the root mean square (r.m.s) of the concentration, $T_Y$ is the memory time and $\xi$ a Gaussian with noise : $\langle \xi(t)\xi(t') \rangle = \delta(t-t')$. 
%Note that $T_Y$ is considered to be larger than the typical motor switching time.
%
Therefore $\delta X$ is a normal Gaussian process with zero mean and a r.m.s equal to 1. 
The phosphorylation reaction being thermally activated with a barrier height depending in first approximation, linearly with CheY-P concentration, the switching time for the transition CCW $\to$ CW then reads:
\begin{equation}
\begin{aligned}
\tau_s &=\tau_0.\text{exp}(-\Delta_n \delta X)
\end{aligned}
\end{equation}
with $\tau_0$ the typical switching time corresponding to the mean concentration $Y_0$ and $\Delta_n$ measuring the switching time sensitivity to variation in $\lbrack Y \rbrack$.
The $\delta X$ distribution being Gaussian, one should essentially obtain a log-normal distribution for the CCW $\to$ CW transition rates. 
Therefore, at low values of CheY-P, a bacterium undergoes long runs whereas at high value of CheY-P tumbling events are frequent. A bacterium will slowly change its behaviour only after times typically larger than the memory time $T_Y$. In the simulations, the parameter values for the run and tumble dynamics are taken from Figueroa \etal \cite{figueroa20203d}, i.e  $\Delta_n=1.62$, $\tau_0=\exp(1.53)\SI{}{\second}$, $T_Y=\SI{19}{\second}$.\\

The tumbling phase undergone by multi-flagellated bacteria as E.coli is a complex process both hydrodynamically and also biochemically. During the tumbling phase, the switch to CW rotation does not necessarily concerns all motors at the same time. Phenomenologically, one can observe debundling associated with directional changes while bacteria are still swimming, or even observe a debundled flagellum with no change of the swimming direction \cite{mears2014escherichia}. From the two colors experiments in this report, the unbundling phase was defined as the time for which at least one flagellum can be observed out of the bundle. With this definition the mean unbundling duration, was found to be $\langle \tau_{un} \rangle=0.8s$.
%%%

 \begin{figure}[ht!]
\centering
	\includegraphics[width= 7cm]{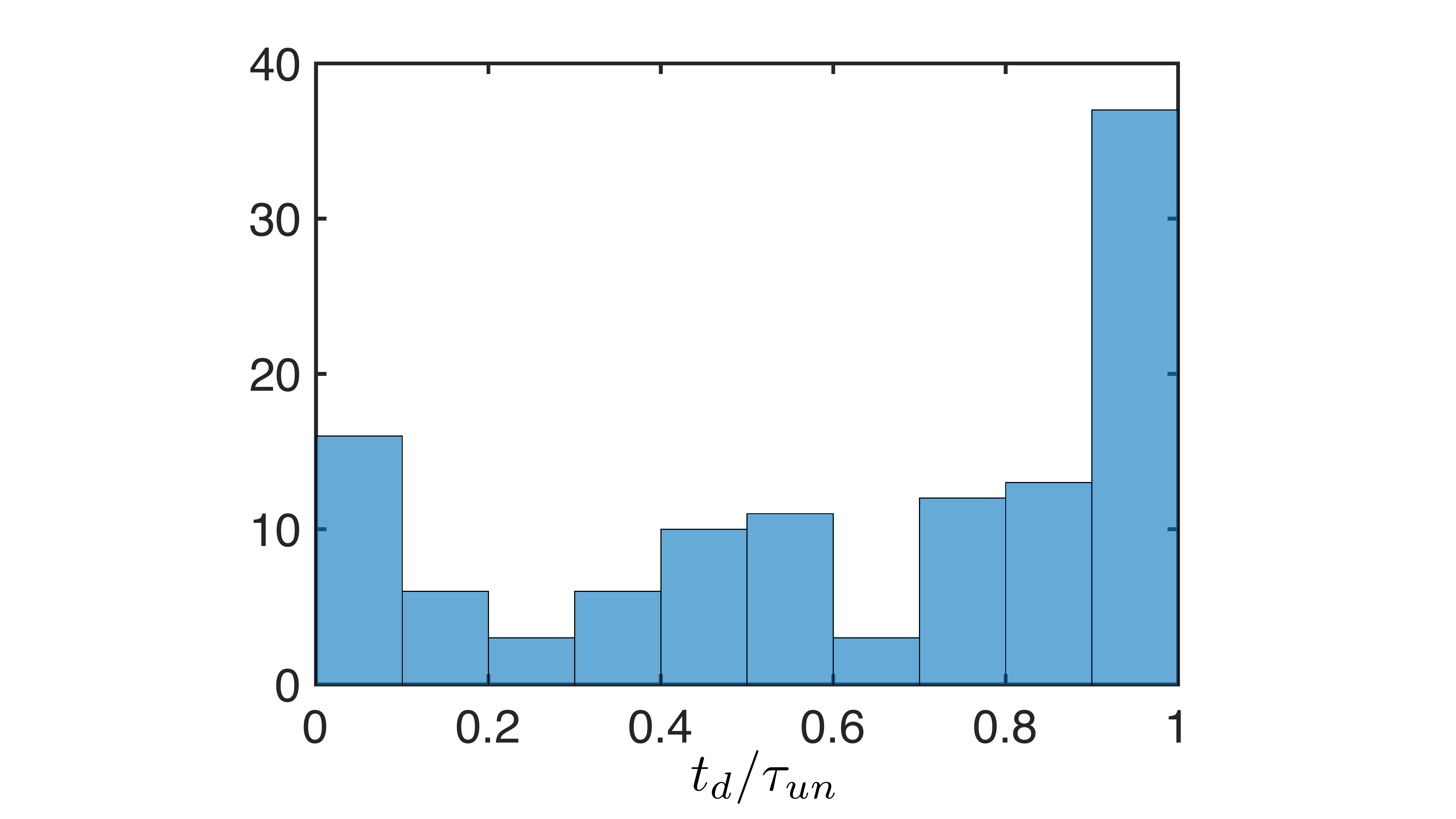}
	\vspace{0 mm}
	\caption{Histogram of the fraction of time $\alpha=t_d/\tau_{un}$, with $\tau_{un}$ the unbundled time and $t_d$ the time during which the bacterium is disorientated during $\tau_{un}$.}
	\label{fraction_tumble_desorientation}
\end{figure}

Here, we do not want to go into too fine details of description as essentially the focus of the study is on long time processes. Following Saragosti et al. \cite{saragosti2012modeling}, we model changes of direction as a rotational diffusion, with a diffusion coefficient $D_r$, lasting a time $\tau_{t}$. Note that what really counts to assess directional changes, is the adimensional diffusion coefficient $\tilde{D_r}= D_r t_{tumb} $, with $t_{tumb}$ the characteristic tumbling time. To complete the kinematic description, we need to obtain $t_{tumb}$ during which the diffusion process applies and in an effective way would essentially correspond to motors in the CW rotation phase.
Experimentally, by monitoring the variations of orientations during unbundling events, we found that indeed bacteria do not fully disorient during the whole unbundling duration but only during a fraction of it we call $\alpha$.
We associate the disorientation of the bacterium with an increase of the absolute value of its rotational velocity $v_{\theta} = |\bo{\dot{p}}|$. During a unbundling event of duration $\tau_{un}$, we compute the time $t_d$ during which $v_{\theta}$ is greater than a threshold value $v_m = \langle v_{\theta} \rangle$, which is the mean value of the velocity over all the trajectories. We then compute the fraction of time $\alpha = t_d/\tau_{un}$ which is the fraction of unbundled time during which the bacterium is disoriented. Note that $\alpha$ will eventually depends on the threshold value. All we want to say here is that a bacterium is not disoriented during all the unbundling event.
In Fig. \ref{fraction_tumble_desorientation}, we display the histogram of $\alpha$. Even though there seems to be a slight clustering of events around $\alpha = 1$ we choose to simulate the distribution of $\tau_{t}$ as a Poisson process of mean time $t_{tumb} = \langle \alpha \tau_{un} \rangle = \SI{0.4}{\second}$, where  $\alpha$ is a random variable chosen between $0$ and $1$ prior to each $CCW \rightarrow CC $ (i.e. run to tumble) switch. Past this time, the bacterium ``effective motor`` will resume in the CCW (run) phase. As obtained experimentally by Figueroa \etal \cite{figueroa20203d}, we use $\tilde{D_r}= 3.86$, to simulate the rotational diffusive process.\\

\emph{-~Description of the simulated trajectories in the bulk} \\
A bacterium swims between two planes, located at $z=0$ and $z=h$. Here we only describe the part 
corresponding to the bulk (i.e $z\ne 0$ or $h$) which was already introduced by Figueroa \etal \cite{figueroa20203d}. Kinematics at the surfaces is described in the main part of the text. In the run phase, a bacterium swims at a constant velocity $v_b=\SI{26}{\micro\meter\per\second}$ in the direction of its orientation vector $\bo{p}$. The evolution of its position and orientation follow the equations: 
\begin{equation} \label{kinematics_positions_run}
\begin{gathered}
\begin{aligned}
\dot{\bo{r}} &= v_b \bo{p} \\
\dot{\bo{p}} &=0
\end{aligned}
\end{gathered}
\end{equation}
Then, during the tumbling phase, the bacterium stops and its orientation $\bo{p}$ undergoes a reorientation process lasting a time $\tau_t$ modelled by a diffusion on the unit sphere with a rotational diffusion coefficient $D_r = \tilde{D_r}/t_{tumb}$. After time $\tau_t$ , it will resume in the run phase.\\

\emph{-~Comparaison between the BV model and Poisson model}

In a previous article from our group \cite{figueroa20203d} we showed that the classic run time distribution, introduced by Berg \cite{Berg1972}, failed to reproduce our experimental data and that one needs to introduce the BV model to explain the experimental findings. 

\begin{figure}[ht!]
\centering
	\includegraphics[width= 7cm]{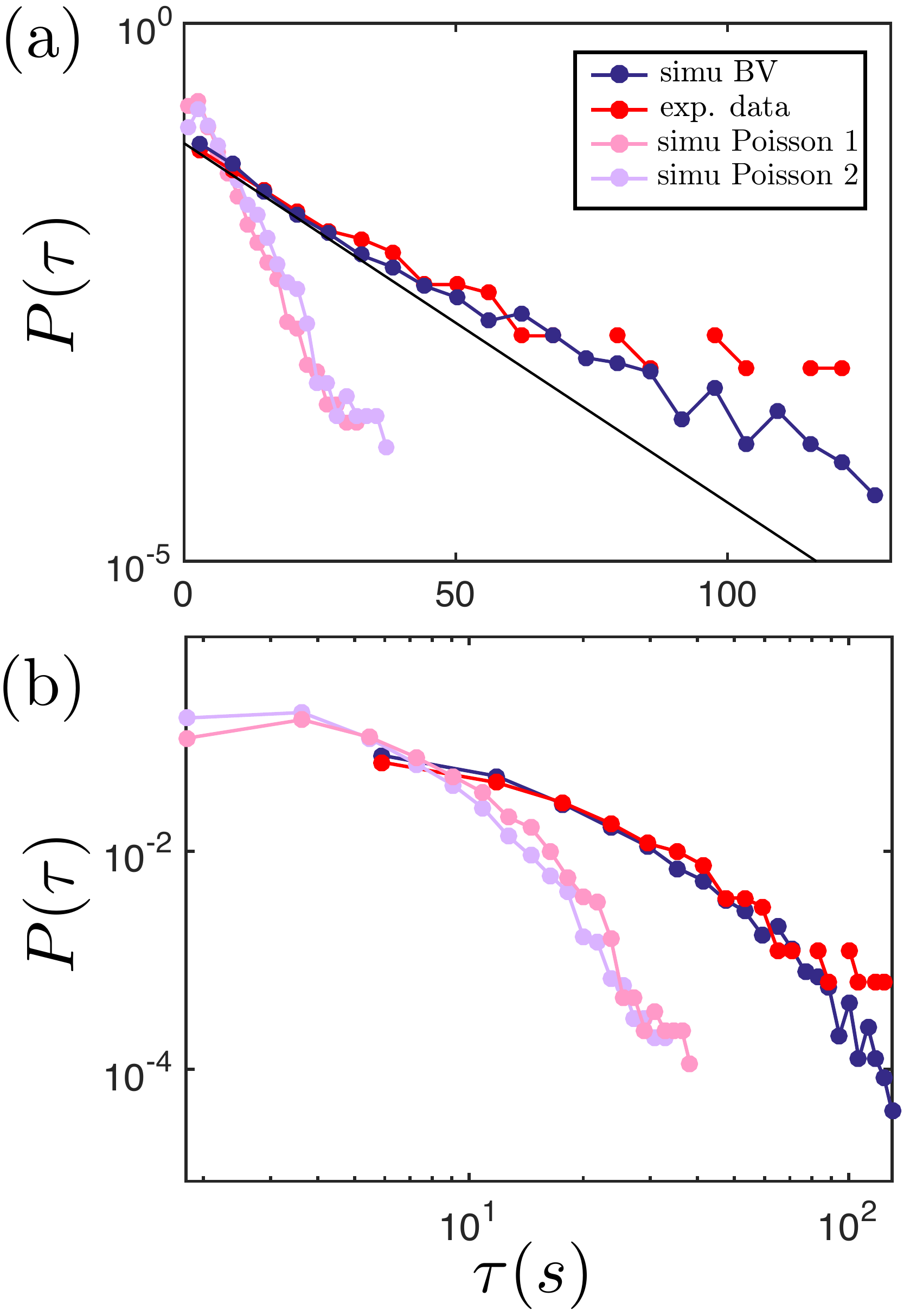}
	\vspace{0 mm}
	\caption{Residence time distribution in (a) log/lin and (b) log/log scale comparing the BV model to the standard Poisson model ($\langle \tau_r \rangle=\SI{1}{\second}$), simu Poisson 1 is obtained with the standard tumbling time $\langle \tau_t \rangle=\SI{0.1}{\second}$ while simu Poisson 2 with $\langle \tau_t \rangle=\SI{0.4}{\second}$. The black line in (a) is an exponential decay of characteristic time scale \SI{14}{\second}. }
	\label{SI_PDF_tau_log_lin}
\end{figure}

Figure.~\ref{SI_PDF_tau_log_lin} displays the residence time distribution in semi-log and log-log scale for the experiment and different simulations. Two types of simulations have been performed : a simulation using the BV model (same data as in Fig.~2 in the manuscript) and two simulations using a Poisson distribution for the run time. Beside that, the simulations have been performed the same way.  
As one can see, a Poisson distribution for the run time does not reproduce the experimental data and changing the typical tumbling time has a minor effect on the distribution. Simulation using the BV model reproduce well the experimental outcome. This strengthened our claim that one needs to consider the behavioral variability in the run time to account for the run time distribution as well as for the residence time distribution.

\begin{figure}[ht!]
\centering
	\includegraphics[width= 7cm]{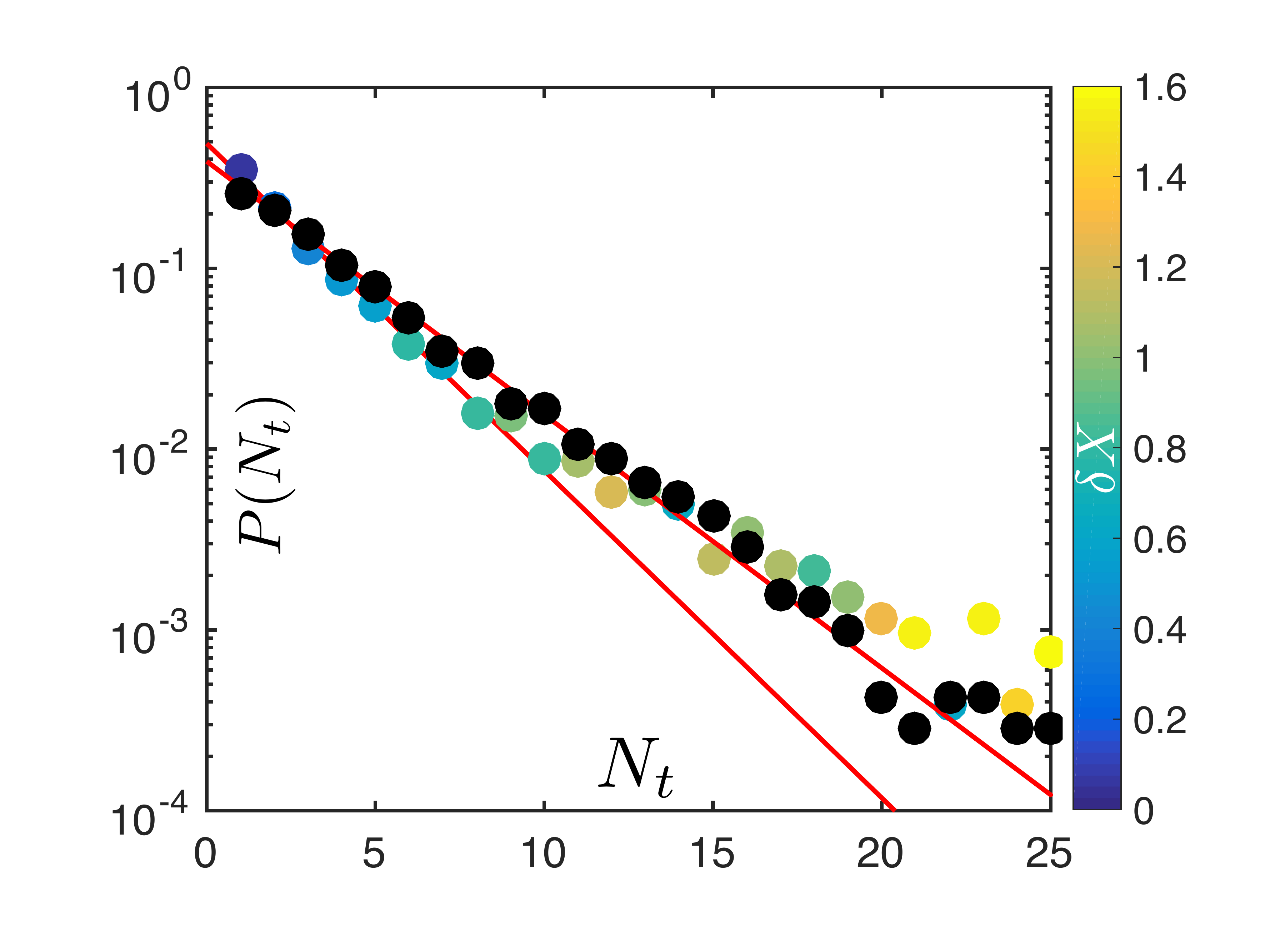}
	\vspace{0 mm}
	\caption{Distribution of number of tumbles $N_t$ before escape in simulations using the BV model (colored symbol) and Poisson model with $\langle \tau_t \rangle = \SI{0.1}{\second}$ (black symbols). The red lines are proportional to $P(N_t) = (1-p)^{N_t -1}p$. For the BV model: $p = \frac{1}{2.9}$, which gives a mean number of tumbles of $\langle N_{t} \rangle^\ast=2.4$, compared to the mean of the measured distribution $\langle N_t \rangle = 3.5$. For the Poisson model: $p = \frac{1}{3.6}$ which gives $\langle N_{t} \rangle^\ast=3.1$. Symbol colors show the corresponding dimensionless average CheY-P concentration $\delta X$ when leaving the surface (only present in the BV model).}
	\label{Chey_Berg_comparaison}
\end{figure}

In Figure.~\ref{Chey_Berg_comparaison}, we also compare the distribution of the number of tumbles needed to escape in simulation using the BV model and the standard Poisson model. As previously commented in the manuscript, the distribution coming from the BV model exhibits two regimes, accounting for the two sub-populations in the close surface region. The transition occurs around $N_t \sim 10$ for a $\delta X \sim 1$ corresponding to a typical run time of $\sim \SI{1}{\second}$ (i.e the typical run time of the standard Poisson model). The distribution coming from the standard Poisson model is a simple exponential decay with a slope that is somehow an average between the two slopes of the BV distribution.

\bibliographystyle{apsrev4-1}

\bibliography{Bibliography_article_suface}

%References [35-40] are in the Supplemental Material not already in paper.

%
\end{document}